\documentclass[10pt]{article}

\usepackage{amsmath}
\usepackage{amssymb}

\usepackage{graphicx}

\usepackage{cite}

\usepackage{color}

\usepackage[labelfont=bf,labelsep=period,justification=raggedright]{caption}

\makeatletter
\renewcommand{\@biblabel}[1]{\quad#1.}
\makeatother

\date{}

\begin{document}

% Title must be 150 characters or less
\begin{flushleft}
{\Large
\textbf{Roles of Ties in Spreading}
}
% Insert Author names, affiliations and corresponding author email.
\\
Ai-Xiang Cui,
Zimo Yang,
Tao Zhou$^{\ast}$
\\
Web Sciences Center, University of Electronic Science and Technology of China, Chengdu 610054, People's Republic of China
\\
$\ast$ E-mail: zhutou@ustc.edu
\end{flushleft}

% Please keep the abstract between 250 and 300 words
\section*{Abstract}

\emph{Background}: Controlling global epidemics in the real world and accelerating information propagation in the artificial world are of great significance, which have activated an upsurge in the studies on networked spreading dynamics. Lots of efforts have been made to understand the impacts of macroscopic statistics (e.g., degree distribution and average distance) and mesoscopic structures (e.g., communities and rich clubs) on spreading processes while the microscopic elements are less concerned. In particular, roles of ties are not yet clear to the academic community.

\emph{Methodology/Principle Findings}: Every edges is stamped by its strength that is defined solely based on the local topology. According to a weighted susceptible-infected-susceptible model, the steady-state infected density and spreading speed are respectively optimized by adjusting the relationship between edge's strength and spreading ability. Experiments on six real networks show that the infected density is increased when strong ties are favored in the spreading, while the speed is enhanced when weak ties are favored. Significance of these findings is further demonstrated by comparing with a null model.

\emph{Conclusions/Significance}: Experimental results indicate that strong and weak ties play distinguishable roles in spreading dynamics: the former enlarge the infected density while the latter fasten the process. The proposed method provides a quantitative way to reveal the qualitatively different roles of ties, which could find applications in analyzing many networked dynamical processes with multiple performance indices, such as synchronizability and converging time in synchronization and throughput and delivering time in transportation.

%\section*{Author Summary}

\section*{Introduction}
Early before the classification of social ties proposed, in 1954, the Russian mathematical psychologist Anatol Rapoport \cite{rapoport1954spread} has been aware of ``well-known fact that the likely contacts of two individuals who are closely acquainted tend to be more overlapping than those of two arbitrarily selected individuals." This argument became one of the cornerstones of social network theory. In 1973, ties in social networks, generally, come in two varieties: strong and weak, which has been first proposed by the American sociologist Mark Granovetter in Ref. \cite{granovetter1973strength}, one of the most influential sociology papers ever published. Different relationships can be measured in the currency of tie strength. According to the closeness, connections to close friends have been said to be ``strong'' ties, while those to acquaintances have been called ``weak'' ties \cite{granovetter1973strength,granovetter1995getting,erickson1978flow,lin1986access,shi2007networks}. Tie strength usually plays an vital role in many real networks and is crucial to understand dynamical processes on the networks \cite{shi2007networks,zhao2010weak,brown1987social}.  Strong ties are the people you really trust, people whose social circles tightly overlap with your own. Often, they are also the people most like you. Weak ties, conversely, are merely acquaintances and often provide access to novel information \cite{shi2007networks}. Weak ties display an important bridging function \cite{zhao2010weak}, while strong ties are more likely to activated for the flow of referral information and more influential than weak ties \cite{brown1987social}. In addition, weak ties could play a more significant role than strong ties to keep the stability \cite{csermely2006weak}, maintain the connectivity \cite{onnela2007structure} and uncover the missing information \cite{lu2010link}.

In despite of the qualitative distinction between strong and weak ties, the tie strength could be quantitatively described by edge weight---the edges with high weights are considered to be strong. In a number of social networks, edges are often associated with weights that differentiate them in terms of their strength, intensity, capacity or the frequency of recent contacts \cite{granovetter1973strength,barrat2004architecture}. For non-social networks, weights often refer to the functions performed by edges, e.g., the amount of traffic flowing along connections in world-wide airport networks \cite{barrat2004architecture}, the number of joint papers of two coauthors in scientist collaboration networks \cite{barrat2004architecture}, the number of synapses and gap junctions in neural networks \cite{rojas1996neural}, the carbon flow between species in food webs\cite{nordlund2007identifying}. Yan \textit{et al.} \cite{gang2005epidemic} investigated the epidemic spreading in weighted scale-free networks and the simulation results indicated that the more homogeneous weight distribution of the network, the more quickly epidemic spreads on it. This finding was further demonstrated by an edge-based mean-field solution \cite{zimo2011epidemic}. Chu \textit{et al.} \cite{chu2011epidemic} showed that the weight distribution has strong impacts on both epidemic threshold and prevalence. Baronchelli and Pastor-Satorras \cite{baronchelli2010mean} considered the diffusive dynamics on weighted networks and shed light on the validity of mean-field theory on weighted networks. These studies revealed quantitative impacts of edges with different weights, but ignored the qualitatively different roles respectively played by strong and weak ties.

In this paper, we propose a quantitative method to uncover the qualitatively different roles of strong and weak ties via a weighted susceptible-infected-susceptible (SIS) model. Experimental results on six real networks emphasize the different roles of strong and weak ties: the former improve the steady-state fraction of infected nodes while the latter fasten the spreading speed.

% Results and Discussion can be combined.
\section*{Results}

\subsection*{Model}
The SIS model\cite{baily1975mathematical} is suitable to describe the cases when the individuals cannot acquire immunity after recovering from the disease, such as influenza, pulmonary tuberculosis and gonorrhea. With disease of this kind individuals that are cured usually catch again. In the networked SIS model, nodes are in two discrete states, ``susceptible'' or ``infected''. Each infected node will contact all its neighbors once at each time step, and therefore the infectivity of each node is proportional to its degree. In the real world, individuals may be only able to contact limited population within one time step\cite{zhou2006behaviors}. For example, salesman in network marketing processes will not make referrals to all his acquaintances due to the limited money and time\cite{kim2006network}. In sexual contact networks, although a few individuals have hundreds of sexual partners, their sexual activities are not far beyond a normal level due to the physiological limitations\cite{liljeros2003sexual,schneeberger2004scale}. Therefore, in the present model we assume every individual has the same infectivity \cite{yang2007epidemic,zhou2006behaviors}. Without the lose of generality, at each time step, each infected node will select one of its neighbors to contact. If the selected neighbor has been infected already, nothing happens, while if it is susceptible, with probability $\alpha$, it will be infected. Meanwhile, each infected node will become susceptible in the next time step with probability $\beta$. The probability an infected node $i$ will select its neighbor $j$ is
\begin{equation}
p_{ij}=\frac{s^{b}_{ij}}{\sum_{l\in{\Gamma_{i}}}s^{b}_{il}},
\end{equation}
where $\Gamma_{i}$ is the neighbor set of node $i$, $s_{ij}$ denotes the tie strength between $i$ and $j$, and $b$ is a tunable parameter. If $b=0$, the infected node randomly selects a neighbor to contact. If $b>0$, strong ties are favored to constitute the paths of spreading, while if $b<0$, weak ties are favored.

In different contexts, the strength of a tie may have different definitions and measures\cite{marsden1984measuring,lue2011link}, which may depend on external information to network topology. For general networks, one may be not aware of external information and thus it is meaningful to give a natural definition solely based on network topology. According to Rapoport's theory\cite{rapoport1954spread} and other supportive observations\cite{kossinets2006empirical,liben2007link,zhou2009predicting} and models\cite{holme2002growing,cui2011emergence}, we define the tie strength $s_{ij}$ in spite of $i$ and $j$'s common neighbors, as follows:
\begin{equation}\label{sij}
s_{ij}=\frac{n_{ij}+\delta}{\sqrt{k_{i}k_{j}}},
\end{equation}
where $n_{ij}$ is the number of common neighbors of $i$ and $j$, $k_i$ is the degree of $i$, and $\delta$ is a constant that gives chance to the tie connecting two nodes without common neighbors. For simplicity, we set $\delta=1$. The dynamical process starts with randomly selecting a certain number of infected nodes.

\subsection*{Data}
To see the impacts of tie strength on the spreading dynamics, experiments are carried out on six real networks: (1) Facebook-like Social Network (FSN) \cite{opsahl2009clustering}: it originates from an online community for students at University of California. The dataset includes the users that have sent or received at least one message, and an undirected edge is set between $i$ and $j$ if $i$ has sent (received) an online message to (from) $j$. (2) Enron Email Network (EEN) \cite{leskovec2009community,klimt2004introducing}: it covers about half million emails. Nodes of this network are email addresses and if an address $i$ sent at least one email to address $j$, an undirected edge is established between $i$ and $j$. (3) Slashdot Social Network (SSN) \cite{leskovec2010signed}: nodes in this network are the users in Slashdot, which is a technology-related news website, and edges represent friendships between users. (4) Epinions Social Network (ESN) \cite{richardson2003trust}: this is a who-trust-whom online social network. Nodes are the members of the general consumer review site \emph{Epinions.com}, and edges represent the trust relationships between two members. (5) Gnutella Peer-to-peer Network (GPN) \cite{ripeanu2002mapping}: Nodes represent hosts and edges stand for connections between the hosts. (6) Oregon Autonomous Systems (OAS) \cite{leskovec2005graphs}: this is an AS-level Internet topology graph obtained by the \emph{Route Views Project}. In order to guarantee the connectivity, we use the largest connected components of these networks. Table \ref{basic-characteristic} presents the basic statistics of the largest connected components of the six networks.

\subsection*{Results and Analysis}
\begin{figure}[!ht]
\begin{center}
\includegraphics[width=4in]{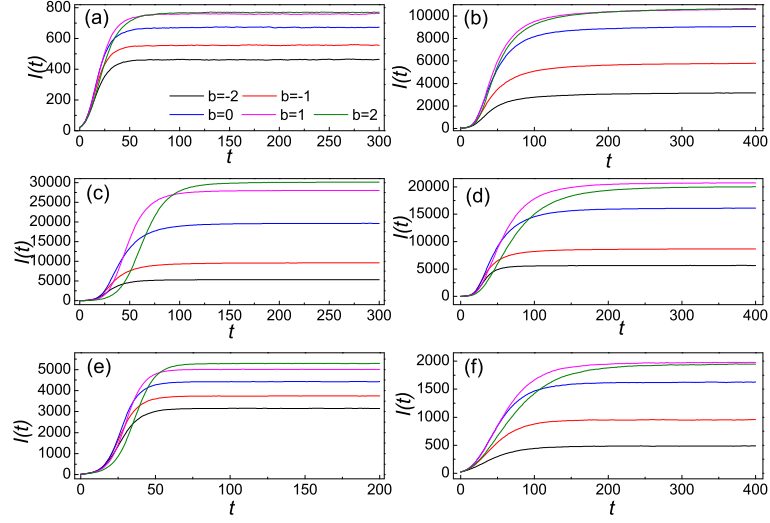}
\end{center}
\caption{\textbf{Spreading processes for different values of the parameter $b$.} These six plots show how the number of infected individuals, $I(t)$, changes with time on the six real networks: (a) Facebook-like social network, (b) Enron email network, (c) Slashdot social network, (d) Epinions social network, (e) Gnutells peer-to-peer network, (f) Oregon autonomous systems. Results are obtained by averaging over 1000 independent realizations.}
\label{spread-process}
\end{figure}

All simulations start with 20 randomly selected nodes as infected sources, with infectivity rate $\alpha=0.4$ and recovery rate $\beta=0.1$. Figure~\ref{spread-process} gives us intuition of the spreading processes following the proposed model, in which one can obverse the number of infected nodes, $I(t)$, changing with time, $t$, and considerable effects made by $b$. We next investigate $b$'s impacts on two macroscopic features of the spreading process---the prevalence and speed. The first one is characterized by the fraction of infected nodes in the steady state, $\rho=I(\infty)/N$. For the second one, if we define $\Delta I(t):=I(\infty)-I(t)$, it converges to zero following an exponential form (it is easy to be checked numerically and thus not shown here), $\Delta I(t) \propto e^{-\lambda t}$, and the parameter $\lambda$ is an informative index for the converging speed. The larger $\rho$ the wider the spreading, and the larger $\lambda$ the faster the spreading.

\begin{figure}[!ht]
\begin{center}
\includegraphics[width=4in]{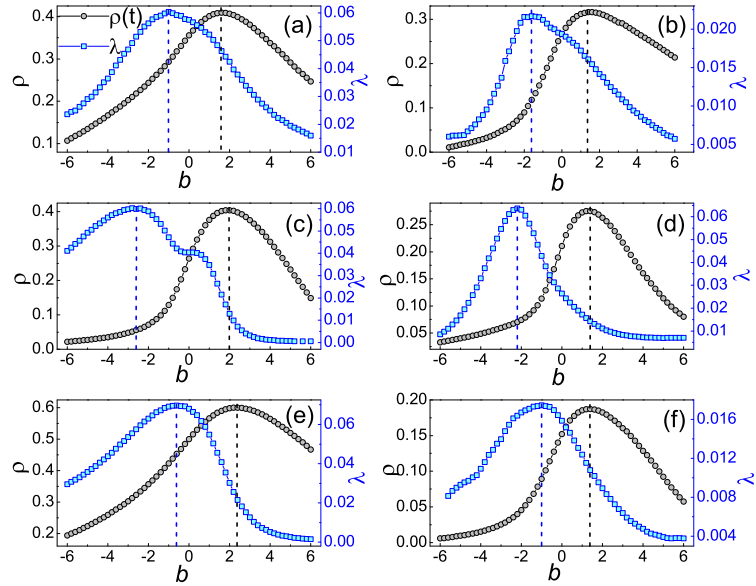}
\end{center}
\caption{\textbf{Steady-state infected density $\rho$ and speed index $\lambda$ for different $b$.} The serial numbers (a)-(f) have the same meaning to those for figure 1. The black and blue curves represent the steady-state infected density and the speed index, respectively. The black and blue dash lines indicate the peaks of the corresponding curves. Results are obtained by averaging over 1000 independent realizations.}
\label{density-and-exponent}
\end{figure}

Figure~\ref{density-and-exponent} displays how the steady-state infection density $\rho$ and the speed index $\lambda$ change with $b$ based on the six real networks. For each case, there are well-defined peaks for both $\rho(b)$ and $\lambda(b)$. More interestingly, in every case, all the ``optimal" values of $b$ subject to high prevalence are larger than zero, while all the ``optimal" values of $b$ subject to high speed are smaller than zero. These phenomena indicate the different roles of strong and weak ties: the former operate locally and keep a high prevalence in local surrounding while the latter bridge different densely connected groups and enhance the chance to infect healthy groups. As an imprecise metaphor to global epidemic spreading, the strong ties are similar to the urban buses and gathering places while the weak ties are similar to the intercity trains and airlines.

\begin{figure}[!ht]
\begin{center}
\includegraphics[width=4in]{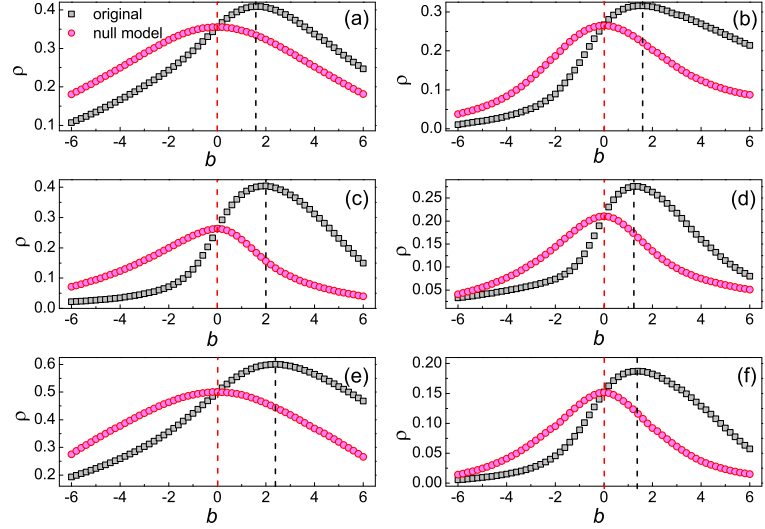}
\end{center}
\caption{\textbf{Comparing $\rho(b)$ between real and randomized networks.} The serial numbers (a)-(f) have the same meaning to those for figure 1. The black and red curves/lines represent the cases for real and randomized networks, respectively. Results are obtained by averaging over 1000 independent realizations.}
\label{density-compare}
\end{figure}

\begin{figure}[!ht]
\begin{center}
\includegraphics[width=4in]{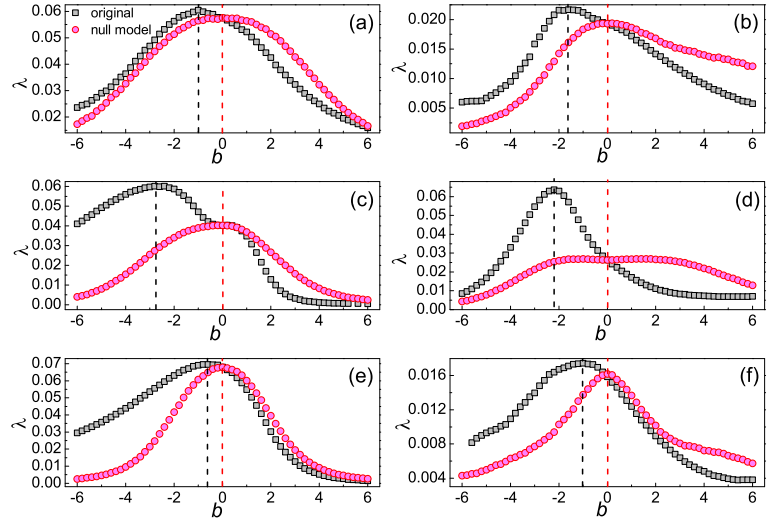}
\end{center}
\caption{\textbf{Effects of correlations between tie strength and network structure on the speed to reach the steady state.} The serial numbers (a)-(f) have the same meaning to those for figure 1. The black and red curves/lines represent the cases for real and randomized networks, respectively. Results are obtained by averaging over 1000 independent realizations.}
\label{exponent-compare}
\end{figure}

To emphasize the significance of our findings, we further consider the null model, where strengths are completely redistributed. This randomization operation destroys the correlation between structure and strength. Figure~\ref{density-compare} reports $\rho(b)$ for real networks and the corresponding randomized networks. Different from the positive peaks for real networks, all peaks for randomized networks are at $b=0$. Figure~\ref{exponent-compare} shows $\lambda(b)$ for real networks and the corresponding randomized networks, again, peaks for randomized networks are all at $b=0$. The results are in accordance with the known conclusion \cite{zimo2011epidemic} that in the absence of correlation between structure and weight, the most homogeneous weight distribution leads to the widest and fastest epidemic spreading. Comparison between results from real and randomized cases suggests that the non-zero ``optimal" values of the parameter $b$ do reveal some significant features of reality.

\section*{Discussion}

We studied a weighted SIS model where an edge's weight is functionally dependent on its strength. Motivated by the \emph{transitivity} in social network analysis \cite{wassermann1994social} and the Rapoport's theory \cite{rapoport1954spread}, we apply a common-neighborhood-based similarity index to quantify edge strength. By comparing the experimental results on six real networks and their counterpart randomized networks, we show that strong and weak ties play distinguishable roles in spreading dynamics: the former enlarge the infected density while the latter fasten the process.

Note that, the optimal values of $b$ subject to the infected density and spreading speed are far different, indicating that the ``optimization" of an spreading process could have multiple performance indices and these indices may conflict to each other. Analogous examples are numerous: bypassing main intersections and arterial roads could enhance the network throughput while increase the delivering time especially in uncongested states \cite{yan2006efficient}, and removing directed loops could enhance the synchronizability while slow down the synchronizing process \cite{zhou2010synchronization}. Our method provides a quantitative way to reveal the qualitatively different roles of ties (also can be extended to deal with nodes) in such dynamical processes.

We hope this work could contribute to the long journey towards fully understanding the relations between structural features and functional performances. Along this way, more experiments on disparate networks will benefit us, and we will look into benchmark network models to see whether  some structural properties, such as the presence of degree-degree correlation and community structure, are responsible for the observations reported here.

% You may title this section "Methods" or "Models".
% "Models" is not a valid title for PLoS ONE authors. However, PLoS ONE
% authors may use "Analysis"
%\section*{Materials and Methods}

% Do NOT remove this, even if you are not including acknowledgments
\section*{Acknowledgments}
We acknowledge Pak-Ming Hui for valuable discussions. This work is partially supported by the National Natural Science Foundation of China under grant number 90924011, and the Fundamental Research Funds for the Central Universities.

%\section*{References}
% The bibtex filename
\bibliography{reference}

\section*{Figure Legends}

Fig.\ref{spread-process} \textbf{Spreading processes for different values of the parameter $b$.}

Fig.\ref{density-and-exponent} \textbf{Steady-state infected density $\rho$ and speed index $\lambda$ for different $b$.}

Fig.\ref{density-compare} \textbf{Comparing $\rho(b)$ between real and randomized networks.}

Fig.\ref{exponent-compare} \textbf{Comparing $\lambda(b)$ between real and randomized networks.}

\section*{Tables}

\begin{table}[!ht]
\centering
\caption{
\bf{Basic characteristics of the largest connected components,} where $N$ and $E$ represent the number of nodes and edges, respectively.}
\begin{tabular}{lllllll}
\hline
Dataset &FSN &EEN &SSN &ESN &GPN &OAS\\
\hline
\textit{N} &1,893 &33,696 &74,444 &75,868 &8,842 &10,670\\
\textit{E} &13,835 &180,811 &382,456  &405,729 &31,837 &22,002\\
\hline
\end{tabular}
\begin{flushleft}
\end{flushleft}
\label{basic-characteristic}
\end{table}

%\newpage

\end{document}